# Reanalyzing variable directionality of gene expression in transgenerational epigenetic inheritance

Abhay Sharma

CSIR-Institute of Genomics and Integrative Biology

Council of Scientific and Industrial Research

Sukhdev Vihar, Mathura Road

New Delhi 110025

Telephone: +91-11-26932421

Fax: +91-11-27662407

Email: abhaysharma@igib.res.in

**Abstract**:

A previous report claimed no evidence of transgenerational epigenetic inheritance in a mouse model of *in utero* environmental exposure, based on the observation that gene expression changes observed in the germ cells of G1 and G2 male fetus were not in the same direction. A subsequent data reanalysis however showed a statistically significant overlap between G1 and G2 genes irrespective of direction, leading to the suggestion that, as phenotypic variability in epigenetic transmission has been observed in several other examples also, the above report provided evidence in favor of, not against, transgenerational inheritance. This criticism has recently been questioned. Here, it is shown that the questions raised are based not only on incorrect statistical calculations but also on wrong premise that gene expression changes do not constitute a phenotype.

*Introduction*

Iqbal *et al.* [1] previously claimed, based mainly on gene expression data, that endocrine disrupting chemicals (EDCs) do not cause transgenerational effects in mammals. In brief, the authors treated G0 female mice with the EDCs vinclozolin (VZ), and di-(2-ethylhexyl)phthalate (DEPH), performed transcriptomic analysis of purified G1 and G2 prospermatogonia, found statistically significant overlap neither between upregulated genes in G1 and G2 nor between downregulated genes in the two prospermatogonia samples, and concluded that the EDCs do not cause TEI. My reanalysis of their data [2] using hypergeometric distribution probability showed that the overlap becomes highly significant if both up- and down-regulated genes are combined together, and suggested, citing known examples of phenotypic variability observed across generations in epigenetic inheritance, that an overlap of differentially expressed genes, irrespective of directionality of expression change, supports TEI. A comparison between Iqbal *et al.*'s VZ and DEPH associated differentially expressed genes with previously reported differentially expressed genes in embryonic and adult tissues drawn from multiple generations of rats, in studies investigating transgenerational effects of VZ, showed a highly significant VZ-VZ, not VZ-DEPH, overlap. This further supported occurrence of TEI. Recently, in response to my criticism of Iqbal *et al.*'s claim, Szabó's [3] reanalyzed the gene expression data and claimed that my statistical calculations were wrong. She also asserts that the evidence that I cited in support of observing variable directionality of phenotypic expression across generations are not relevant to gene expression studies. Here, both of her concerns are addressed for clarification.

*Data reanalysis*

Szabó's [3] first objects that I selectively focused on genes identified at reduced statistical stringency by Iqbal *et al.* [1]. It is however notable that Iqbal *et al.* themselves used these genes in Fisher's exact test to arrive at the conclusion that a significantly higher number of the common changes between generations occurred in the opposite direction. This objection is therefore not relevant. On the contrary, Iqbal *et al.*'s above conclusion reiterates the core concern, that why should gene changes in opposite direction be assumed as negative evidence in transgenerational inheritance. Next, Szabó asserts that Iqbal *et al.*'s conclusion, that they did not find any evidence of TEI, was not erroneous, as claimed by me, mainly because she finds my overlap analysis incorrect. In support, Szabó produces two data sets, Tables 1 and 2 [3], countering Figures 1 and 2 [2] of my analysis, respectively. The Table 1 [3] shows the hypergeometric *p* values for significance of overlap between the combined set of up- and down-regulated genes in the first and the second generation, as does my Figure 1 [2]. A comparison of the data published in my Figure 1 [2] and that in Szabó's Table 1 [3] clearly shows (**Figure 1**) that Szabó's result does not dispute my finding, that differentially expressed genes between generations overlap significantly.

The rest of the data shown in Table 1 [3] relate to comparisons between genes that changed in expression either in the same or in the opposite direction across generations, results that are not directly related to my analysis. Nevertheless, these comparisons further support TEI, because a highly significant ($p$ = 8.94E-04) overlap is reported for genes that changed in the opposite direction between generations in VZ group. Had there been no transgenerational effects, a significant overlap would not have been observed. Szabó's explanation that genes changing in opposite direction may indicate a slight overcompensation in the erasure process is not supported by Iqbal *et al.*'s data. Though Iqbal *et al.* performed extensive DNA methylation analysis, they found no evidence of methylation changes across generations. Moreover, should Szabó's speculation stand valid, would it not beg the question as to why overcompensation will be observed if there is no transgenerational effect? Regarding Szabó's objection [3] that adjustment for multiple hypothesis testing was not performed by me [2], it is noted here that the significant *p* values shown in her Table 1 [3] as well as my Figure 1 [2] would all remain highly significant even after Bonferroni correction. For example, for the six hypotheses that have been tested in Table 1, the nominal *p* values 1.16E-06, 4.04E-08, and 8.94E-04 become 6.96E-06, 2.42E-07, and 0.005, all significant, after adjustment. Cumulatively, Szabó's Table 1 supports, not contardicts, my Figure 1 that provided evidence in favor of TEI.

As regards Szabó's Table 2, she wrongly counted the cumulative numbers of differentially expressed genes in 36 studies, and of Iqbal *et al.*'s VZ and DEPH genes common to these studies. She simply added all individual gene counts to arrive at cumulative numbers, without removing the duplicate entries. Second, in calculating hypergeometric probabilities, she considered a population of 21041 mouse genes, the total number of genes in the mouse microarrays Iqbal *et al.* used in their mouse study. However, the other 36 studies with which Iqbal *et al.*'s genes are compared were rat studies that used rat microarrays, with different gene coverage. Clearly, a neutral population should have been used to enable normalization of mouse and rat genes. Human genome was used in my analysis exactly to fulfill this requirement.

Regarding Szabó's criticism that why did my Figure 2, unlike one of my previous, unrelated papers, show significance values only for all the 36 studies combined, and not for each of these studies individually, it is noted here that the previous paper was related to miRNA counts wherein a zero "sample success" was not observed for any individual study, with the number of successes observed always being a positive number, mostly in tens or hundreds. In contrast, data related to Figure 2 frequently suffered from zero or very small sample success (**Table 1**). Given that $p$ value in general is dependent on sample size, significance of gene overlaps is a function of gene counts, and increased overlapping significance with growing gene counts is suggestive of the significance observed being a true signal [4], my analysis reported $p$ values for all the studies combined. The extent of overlap (fold enrichment) was nonetheless provided in the figure for individual studies. As regards adjustment for multiple testing, it was obvious from the figure that the nominal VZ-VZ overlapping significance shown ($p$ = 0.0002) can easily survive an adjustment. Even when the $p$ value shown is adjusted for 38 tests, representing 36 individual and 2 combined studies, using Bonferroni correction, the overlap remains significant ($p$ = 0.007). Regarding accompanying change in VZ-DEPH comparison, the change from significant nominal $p$ (0.01) to insignificant adjusted $p$ (0.38) is an expected one, as the two EDCs are expected to cause different gene expression effects at transcriptome level. Further, even when the probability of drawing the given number of successes at least, not exactly, is calculated, the VZ-VZ overlap shown in my figure remains significant. In this calculation, the nominal $p$ value is obtained as 0.00057, and adjusted $p$ as 0.021. Together, Szabó's Table 2 is found invalid, with my Figure 2 data supporting TEI remaining justified.

*Prior evidence*

Szabó argues that the evidence that I cited in support of the possibility that gene expression changes may show directional variability across generations in transgenerational epigenetic inheritance is inappropriate. Primarily, her concern is that a paper that I referred to showed directional change in gene expression between F1/F2 and F3, not between F1 and F2. She asserts that a lack of phenotype under investigation, primordial germ cell defects, in F3 renders this example unacceptable. Is not altered gene expression a phenotype in itself? Then, is not observing this molecular phenotype in F3 an evidence of TEI? Obviously, the answers to these questions are in the affirmative. Szabó's objection is hence not supported. Her next objection is that another paper that I cited relates to expression of miRNA, not mRNA, and to the worm *C. elegans*, not mammals. How it is that miRNA expression change across generations is acceptable as evidence for TEI, whereas that of mRNA not? Also, is not *C. elegans* an established model of TEI? Moreover, as it is often difficult to cite all relevant papers in an article, the examples that I referred to were only illustrative, not exhaustive. The examples nonetheless conveyed the principal observation, that phenotypes, molecular or otherwise, can vary in directionality across generations in epigenetic inheritance. Given this, Szabó's assertion that there exists no support for directionality change in phenotypic expression in TEI does not hold.

*Conclusion*

Szabó's objections to my criticism of Iqbal *et al.*'s claim are not supported. Her data reanalysis is not accurate Also, her assertion that gene expression alterations do not represent a phenotype is grossly misplaced.

**Abbreviations**:

EDCs, endocrine disrupting chemicals; VZ, vinclozolin; DEPH, di-(2-ethylhexyl)phthalate

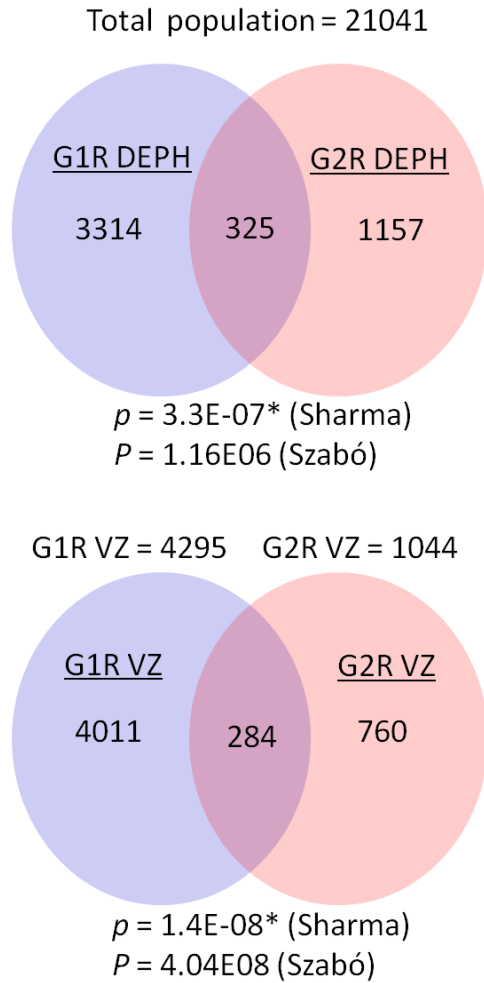

Total population = 21041

G1R DEPH  3314 | 325 | G2R DEPH  1157

*p* = 3.3E-07* (Sharma)
*P* = 1.16E06 (Szabó)

G1R VZ = 4295    G2R VZ = 1044

G1R VZ  4011 | 284 | G2R VZ  760

*p* = 1.4E-08* (Sharma)
*P* = 4.04E08 (Szabó)

**Figure 1**

Comparison of G1R-G2R gene overlap data. Whereas my analysis [2] presented the probability of drawing the given number of successes exactly, Szabó [3] calculated the probability of drawing the given number of successes at least. *indicates a note in the legend of my Figure 1 [2] clearly stating that the *p* value remains highly significant even when the probability is calculated for the given number of successes at least.

**Table 1**. Details of gene overlap analysis presented in Figure 2 [2]

| Previous VZ dataset[*] | Changed genes in dataset[#] | Common with Iqbal et al.'s VZ genes[**,$] | Fold enrichment | Common with Iqbal et al.'s DEPH genes[**,@] | Fold enrichment |
|---|---|---|---|---|---|
| F1 embryo male testis | 49 | 0 | | 0 | |
| F1 embryo E13 male testis | 195 | 5 | 2.3 | 3 | 1.1 |
| F1 embryo E14 male testis | 106 | 2 | 1.7 | 2 | 1.3 |
| F1 embryo E16 male testis | 102 | 3 | 2.7 | 1 | 0.7 |
| F2 embryo male testis | 49 | 0 | | 0 | |
| F3 embryo male testis | 29 | 0 | | 0 | |
| F3 adult male sertoli cells | 285 | 9 | 2.9 | 9 | 2.2 |
| F3 adult female granulosa cells | 326 | 7 | 1.9 | 7 | 1.5 |
| F3 adult female heart | 92 | 0 | | 2 | 1.5 |
| F3 adult female kidney | 337 | 9 | 2.4 | 9 | 1.9 |
| F3 adult female liver | 170 | 0 | | 1 | 0.4 |
| F3 adult female uterus | 176 | 5 | 2.6 | 5 | 2.0 |
| F3 adult male heart | 75 | 1 | 1.2 | 2 | 1.9 |
| F3 adult male kidney | 69 | 0 | | 1 | 1.0 |
| F3 adult male liver | 43 | 2 | 4.3 | 1 | 1.6 |
| F3 adult male prostate | 641 | 12 | 1.7 | 9 | 1.0 |
| F3 adult male seminal vesicle | 139 | 0 | | 0 | |
| F3 adult female ovary | 1939 | 30 | 1.4 | 31 | 1.1 |
| F3 adult male testis | 251 | 3 | 1.1 | 2 | 0.5 |
| F3 adult male amygdala | 137 | 2 | 1.3 | 1 | 0.5 |
| F3 adult male hippocampus | 58 | 1 | 1.6 | 0 | |
| F3 adult female amygdala | 60 | 0 | | 0 | |
| F3 adult female hippocampus | 609 | 13 | 1.9 | 18 | 2.1 |
| F3 adult male brain | 443 | 3 | 0.6 | 4 | 0.6 |
| F3 adult male basolateral amygdala | 10 | 0 | | 1 | 7.2 |
| F3 adult male brain cortex | 39 | 2 | 4.7 | 1 | 1.8 |
| F3 adult male hippocampus CA1 | 29 | 0 | | 0 | |
| F3 adult male hippocampus CA3 | 97 | 1 | 0.9 | 1 | 0.7 |
| F3 adult female basolateral amygdala | 29 | 1 | 3.2 | 0 | |
| F3 adult female brain cortex | 50 | 0 | | 0 | |
| F3 adult female hippocampus CA1 | 18 | 1 | 5.1 | 1 | 4.0 |
| F3 adult female hippocampus CA3 | 12 | 0 | | 0 | |
| F3 adult male ventral prostate | 438 | 5 | 1.1 | 9 | 1.4 |
| F3 adult male prostate epithelial cells | 141 | 2 | 1.3 | 0 | |
| F3 embryo E13 male primordial germ cells | 175 | 3 | 1.6 | 1 | 0.4 |
| F3 embryo E16 male prospermatogonia | 77 | 0 | | 0 | |
| Cumulative (36 studies) | 4066 | 64 | 1.5 | 69 | 1.2 |

*rat study; **mouse study; #within total population, 18865 human genes; $203 genes, within total population; @262 genes, within total population